\documentclass[a4paper,11pt]{article}
\usepackage{subcaption}
\usepackage{comment}
\usepackage{pos}
\usepackage{xcolor}
\usepackage{xargs}
\usepackage{bbm}
\usepackage[colorinlistoftodos,prependcaption,textsize=footnotesize]{todonotes}
\newcommandx{\TODO}[2][1=]{\todo[linecolor=red,backgroundcolor=red!25,bordercolor=red,#1]{TODO: #2}}
\newcommandx{\PH}[2][1=]{\todo[linecolor=green,backgroundcolor=green!25,bordercolor=green,#1]{PAUL: #2}}
\usepackage{layouts}
\usepackage{wrapfig}

\title{Highly anisotropic lattices for Yang-Mills theory}

\author[a]{Kirill Boguslavski}
\author*[a]{Paul Hotzy}
\author[a]{David I.\ M\"uller}
\author[b]{D\'enes Sexty}

\affiliation[a]{Institute of Theoretical Physics, TU Wien,\\
  Wiedner Hauptstraße 8-10, 1040 Vienna, Austria}

\affiliation[b]{Institute of Physics, NAWI Graz, University of Graz,\\
Universitätsplatz 5, 8010 Graz, Austria}

\emailAdd{kirill.boguslavski@tuwien.ac.at}
\emailAdd{paul.hotzy@tuwien.ac.at}
\emailAdd{dmueller@hep.itp.tuwien.ac.at}
\emailAdd{denes.sexty@uni-graz.at}

\abstract{
In this conference proceeding, we investigate the physical anisotropy in terms of the temporal and spatial lattice spacings in relation to the bare parameters of SU(2) pure gauge theory using Wilson gradient flow. Anisotropic lattices have a wide range of applications, from thermodynamic calculations in QCD to very recent real-time simulations using the complex Langevin method. 
We find an almost linear relationship between the bare and renormalized anisotropy. Using a parametrization that includes nonlinear effects and was earlier proposed for SU(3) theory, we obtain a good description of the coupling dependence of the anisotropy with only two fitting parameters. Our observation of an approximately linear relationship and this parametrization should strongly reduce the computational effort of anisotropic lattice calculations in the future. 
}

\FullConference{The 40th International Symposium on Lattice Field Theory (Lattice 2023)\\
July 31st - August 4th, 2023\\
Fermi National Accelerator Laboratory\\}

\begin{document}
\maketitle

\section{Introduction}

Lattice QCD stands out as one of the most successful methods for making non-perturbative predictions in high-energy physics. This approach aims to describe physical processes by discretizing the continuum theory on a hyper-cubic lattice. Given that QCD is a UV-complete and renormalizable theory, the introduction of a scale becomes imperative to give meaning to this discrete representation, as it allows for the determination of the physical value of the lattice spacing. This procedure is commonly referred to as scale-setting. 

Determining the physical length of the lattice spacing can be achieved by comparing observables evaluated on the lattice with their experimentally determined values. Various methods exist for achieving the goal of introducing a scale, both based on phenomenological and theoretical arguments \cite{Sommer:2014mea}. The major challenge with ``direct'' phenomenological approaches, i.e.~by comparing to experimentally measured quantities, often arises from the fact that these observables suffer from a small signal-to-noise ratio, and typically require fitting procedures.

In these proceedings, we employ a theoretical approach — the Wilson gradient flow, initially proposed in \cite{Luscher:2010iy} and further developed in \cite{Borsanyi:2012zs, Ramos:2015baa} — to establish a scale in 3+1D Euclidean lattice gauge theory for SU(2). Of particular interest in this study is the renormalization of lattice anisotropies, where the temporal (Euclidean) and spatial directions are discretized with different resolutions. To address this, we adopt the methodology introduced in \cite{Borsanyi:2012zr}, which relies on the equipartition of the chromo-electric and -magnetic fields within the confined phase of QCD. This work is mainly motivated by complex Langevin simulations of Yang-Mills theory for real-time observables \cite{Berges:2006xc, Berges:2007nr} where increasing the lattice anisotropy has been shown to systematically improve stability \cite{Boguslavski:2022dee, Boguslavski:2023unu}. This may open the door for calculating real-time observables and the QCD equation of state at finite density, which suffer from severe sign problems and currently impede direct calculations. However, there are various other applications of anisotropic lattices ranging from thermodynamic studies to the calculation of quark and glueball masses \cite{Morningstar:1997ff, deForcrand:2017fky}. Additionally, it has been recently proposed to outsource the scale-setting procedures to classical computers due to the limited number of qubits in quantum simulations \cite{Carena:2021ltu}. The renormalization of the anisotropy is crucial for extracting physical values out of all of these calculations. 

We perform a detailed study of the relationship between the gauge coupling and the bare and renormalized anisotropy in Euclidean SU(2) Yang-Mills theory. To the best of our knowledge, such a study has been missing so far for SU(2). Our results indicate that the renormalized and bare anisotropy exhibit an almost linear relationship. It furthermore can be approximately parameterized using a scheme initially studied for SU(3) using the Sommer method \cite{Klassen:1998ua}. This parametrization is especially useful in achieving the continuum limit, as it may eliminate the need for separate calculations to renormalize the anisotropy, which typically incurs high computational costs.

\section{Gradient flow as a scale setting technique}

We are interested in the determination of a scale and the physical lattice anisotropy of SU(2) lattice gauge theory in thermal equilibrium. To achieve this, we employ the gradient flow technique to set the scale for our Euclidean lattice simulations. In the continuum for gauge fields $A_\mu$, the gradient flow equation reads
\begin{align} \label{eq:gradient_flow}
    \frac{\partial \tilde A_{\mu}}{\partial \theta_f} = \tilde D_\nu \tilde F_{\nu\mu}, \quad \left. \tilde A_\mu(x, \theta_f)\right\vert_{\theta_f=0} = A_\mu(x),
\end{align}
where $\tilde D_\nu$ and $\tilde F_{\nu\mu}$ denote the covariant derivative and the field-strength tensor respectively
\begin{align}
    \tilde D_\nu = \partial_\nu + ig [\tilde A_\nu, ],\quad \tilde F_{\mu\nu} = \partial_\mu \tilde A_\nu - \partial_\nu \tilde A_\mu + [\tilde A_\mu, \tilde A_\nu].
\end{align}
We emphasize that, contrary to common practice in Euclidean settings, our indices for the directions of the fields range from $\mu=0,\dots,D$, with $\mu=0$ corresponding to Euclidean time and $D$ denoting the total spatial dimensions of the theory. The flowed gauge fields $\tilde A_\mu(x, \theta_f)$ are distinguished by a tilde symbol and/or the additional argument representing the flow time $\theta_f$. The effect of the gradient flow was studied in \cite{Luscher:2010iy, Harlander:2016vzb} using perturbation theory for the energy density
\begin{align}
    E(\theta_f) = \frac{1}{4} \tilde F_{\mu\nu}^a \tilde F^{\mu\nu}_a.
\end{align}
This perturbative analysis has shown that the flowed energy density is a renormalized quantity for small flow times where the running coupling is small. Beyond this perturbative regime, lattice simulations are used to study the properties of the gradient flow \cite{Luscher:2010iy}. It was confirmed that the gradient flow effectively renormalizes the gauge configurations.

\subsection{Gradient flow on the lattice}
On the lattice, the gradient flow equation is implemented by approximating the Yang-Mills action by the Wilson plaquette action
\begin{align} \label{eq:wilson_action}
    S_\mathrm{W} =  \frac{\beta}{N_c} \left\{\xi_0 \sum_{x, i} \mathrm{Re} \mathrm{Tr} \left[U_{0i}(x) - 1\right] + \frac{1}{\xi_0} \sum_{x, i<j}   \mathrm{Re} \mathrm{Tr} \left[U_{ij}(x) - 1\right]\right\},
\end{align}
with inverse coupling $\beta=2 N_c / g^2$ and bare anisotropy $\xi_0 = a_s / a_\tau$. The plaquettes $U_{\mu\nu}(x)$ are defined as $1\times 1$ Wilson loops in terms of the link variables $U_\mu(x)$
\begin{align}
    U_{\mu\nu}(x) = U_\mu(x) U_\nu (x+\hat{\mu}) U_{\mu}^\dagger (x+\hat{\nu}) U_{\nu}^\dagger (x), \quad U_\mu(x) \simeq \exp\left[ i g a_\mu A_\mu(x+\hat{\mu}/2) \right].
\end{align}
For these link variables, the gradient flow equation is numerically solved by
\begin{align}
    U_\mu(\theta_f + \Delta \theta_f, x) = \exp\left[i \Delta \theta_f t^a W^a_{\mu}(x, \xi_f)\right] U_\mu(\theta, x), \quad U_\mu(0, x) =  U_\mu(x),
\end{align}
where the drift term $W^a_{\mu}$ is given by
\begin{align} \label{eq:flow_drift}
    W^a_{\mu}(\theta_f, x, \xi_f) = \sum_{x, \vert\nu\vert\neq\mu} \rho_{\mu\vert\nu\vert} \mathcal{P}_A\left[ U_{\nu\mu}(\theta_f, x)\right].
\end{align}
The anisotropy enters via the coefficients $\rho_{0j}=\rho_{ij}=1$, $\rho_{i0}=\xi_f^2$ and $\mathcal{P}_A$ projects onto the trace-zero, anti-hermitian part of the matrices:
\begin{align}
    \mathcal{P}_A(U)\equiv \frac{1}{2} \left( U- U^\dagger - \frac{1}{N_c} \mathrm{Tr}\left(U - U^\dagger\right)\right).
\end{align}

To study the effect of the Wilson flow on observables, we introduce the electric (temporal) and magnetic (spatial) contributions to the energy density
\begin{align}
    E(\theta_f) &= E_{\mathrm{st}}(\theta_f) + E_{\mathrm{ss}}(\theta_f), \\
    E_{\mathrm{st}}(\theta_f) = -\sum_{i} \mathrm{Tr} \left\{\mathcal{P}_A\left[ C_{0j}(\theta_f, x) \right]^2\right\}
    &,\quad
    E_{\mathrm{ss}}(\theta_f) = -\sum_{i<j} \mathrm{Tr}\left\{ \mathcal{P}_A\left[ C_{ij}(\theta_f, x) \right]^2\right\}, 
\end{align}
where $C_{\mu\nu}(\theta_f, x)$ denotes the cloverleaves
\begin{align}
    C_{\mu\nu}(\theta_f, x) = \frac{1}{4} \left[ U_{\mu\nu} + U_{(-\nu)\mu} +  U_{(-\mu)\nu} + U_{(-\mu)(-\nu)}\right]_{(\theta_f, x)}.
\end{align}
Compared to the approximation of the field-strength tensor using plaquettes, the cloverleaf approximation reduces cutoff effects on the observable level. However, such artifacts at finite lattice spacing can also be induced from the flow and the generation of the gauge configurations \cite{DallaBrida:2019wur, Ramos:2015baa}.

\subsection{Determination of a lattice scale}

In this section, we summarize the approach to finding a lattice scale using Wilson gradient flow. For this discussion, we consider a symmetric lattice with $\xi_\mathrm{0} = \xi_f = 1$. From dimensional analysis, it follows that the flow time is a quantity of mass dimension $-2$, while the energy density is of dimension $4$. Consequently, $\theta_f^2 E(\theta_f)$ is a dimensionless quantity that allows us to introduce a physical scale. It was proposed to use the condition
\begin{align}
    \theta_f^2 \left. \left\langle E(\theta_f) \right\rangle \right\vert_{\theta_f=t_0} = A = \text{const}
\end{align}
for pure gauge theory to compare flow times of differently discretized configurations and allow the introduction of the scale $t_0$.
An improved version of the $t_0$-scale was introduced in \cite{Borsanyi:2012zs}
\begin{align}
    \theta_f \left. \left\langle \frac{d}{d\theta_f} \theta_f^2 E(\theta_f) \right\rangle \right\vert_{\theta_f=w_0^2}= A.
\end{align}
This $w_0$-scale is a quantity with mass dimension $-1$ and is less sensitive to scales above the cutoff of $1/\sqrt{8\theta_f}$ imposed by the smearing radius of the flow.

In both cases, for $N_c=3$, the observables are evolved until they reach the critical value $A=0.3$. The rationale behind choosing this value is multifaceted. The primary consideration in selecting $A$ is that scaling violations tend to amplify for small flow times due to less effective smoothing of the gradient flow, leading to greater statistical uncertainties. Additionally, the smoothing radius $\sqrt{8t_0}$ can be interpreted as the scale of the system and becomes comparable to the Sommer scale for SU(3) gauge theory. Larger values of $A$ result in increased numerical costs and may necessitate larger lattice volumes, as the smoothing radius might see the periodicity of the lattice, causing finite-volume effects. For SU(2), similar considerations are taken into account, leading to the common choice of $A=0.1$ \cite{Giudice:2017dor, Hirakida:2018uoy}.

\subsection{Determination of the physical lattice anisotropy}
The discussion above is focused on isotropic lattices. When we impose an anisotropic discretization, we can simply use the magnetic part of the energy density to obtain a reference scale
\begin{align} \label{eq:B_flow_condition}
    \theta_f \left. \left\langle \frac{d}{d\theta_f} \theta_f^2 E_{\mathrm{ss}}(\theta_f) \right\rangle \right\vert_{\theta_f=w_0^2}= \frac{A}{2}.
\end{align}
This relation determines the spatial lattice spacing $a_s$ in units of the scale $w_0$.
In \cite{Borsanyi:2012zr}, it was proposed to use an anisotropic flow with $\xi_f \neq 1$ to determine the physical anisotropy of the lattice. This method relies upon the equipartition relation, which should be satisfied in the confined phase of QCD. We introduce the ratio
\begin{align} \label{eq:flow_ratio}
    R(\xi_f) = \left. \left\langle \frac{d}{d\theta_f} \theta_f^2 E^{(\xi_f)}_{\mathrm{ss}}(\theta_f) \right\rangle \right\vert_{\theta_f=w_0^2} \left/ \xi_f^2 \left. \left\langle \frac{d}{d\theta_f} \theta_f^2 E^{(\xi_f)}_{\mathrm{st}}(\theta_f) \right\rangle \right\vert_{\theta_f=w_0^2}\right.,
\end{align}
where the argument of $R$ and superscripts of the observables indicate that the configurations are flowed with an anisotropy of $\xi_f$. In general, these values may not coincide with $\xi_0$. The flow condition in Eq.\ \eqref{eq:B_flow_condition} determines the duration for which configurations are evolved. Note that also the $w_0$-scale depends on $\xi_0$. 
For the renormalized energy density, both temporal and spatial components should contribute equally. Therefore, the goal is to identify the appropriate value for the flow anisotropy $\xi_f$ that results in $R(\xi_f)=1$. In \cite{Borsanyi:2012zr} it was outlined how to systematically find this value and justified that
\begin{align} \label{eq:equi_anisotropy}
    R(\xi_f)=1 \quad \Rightarrow \quad \xi_f = \xi_{\mathrm{phys}},
\end{align}
where $\xi_{\mathrm{phys}}$ denotes the relation between the physical spatial and temporal lattice spacing. This pins down the lattice spacing in temporal direction in terms of the scale $w_0$.

\begin{figure}[t]
    \centering
    \includegraphics{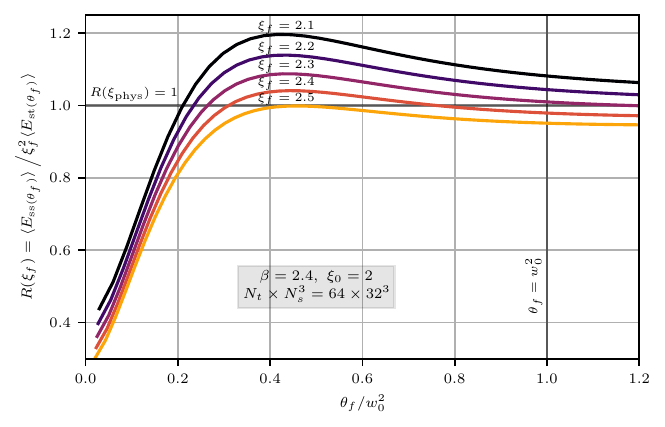}
    \caption{The ratio $R(\xi_f)$ of the magnetic ($E_{\mathrm{ss}}$) and electric ($E_{\mathrm{st}}$) contributions to the energy density is examined for gauge configurations with an inverse gauge coupling of $\beta=2.4$ and a bare anisotropy of $\xi_0=2$. The ratio is shown as a function of flow time $\theta_f$ scaled by $w_0$ for various flow anisotropies $\xi_f$. At $\xi_f=2.3$, the equipartition relation is roughly satisfied at $\theta_f = w_0^2$, indicating a physical anisotropy around this value.
    \label{fig:flow_traj}
    }
\end{figure}

\begin{figure}[t]
    \centering
    \includegraphics[width=0.9\textwidth]{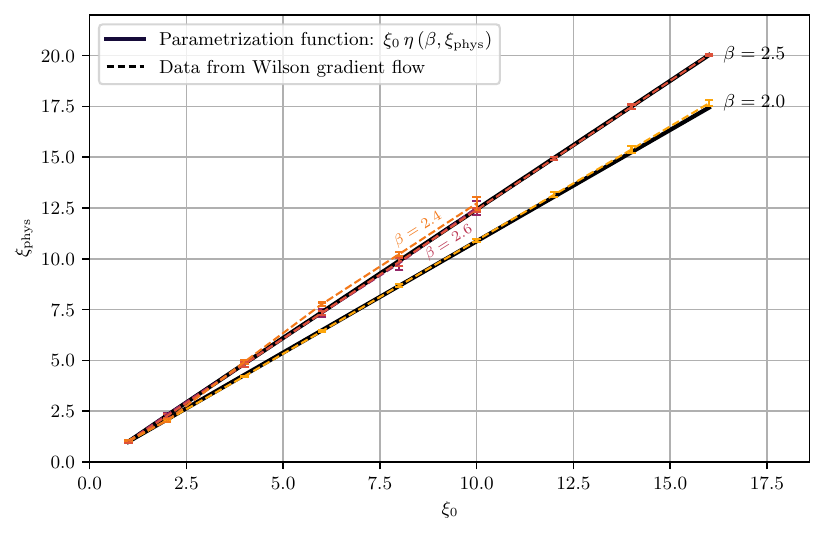}
    \caption{
    Visualization of the physical anisotropy $\xi_\mathrm{phys}$ as a function of the bare anisotropy $\xi_0 \leq 16$ for inverse gauge couplings $\beta = 2.0$ and $2.5$ and with $\xi_0 \leq 10$ for $\beta = 2.4$ and $2.6$. The dashed lines show the data obtained by anisotropic gradient flow while the solid lines show the fit function \eqref{eq:param} for $\beta = 2.0$ and $2.5$.
    \label{fig:anisotropy}
    }
\end{figure}

\section{Numerical results}

We investigate SU(2) pure gauge theory using the anisotropic Wilson action, as described by Eq.\ \eqref{eq:wilson_action}, implemented on a lattice with dimensions $N_t \times N_s^3 = (\xi_0 \cdot 32) \times 32^3$ with anisotropies of $\xi_0=1-16$. Our simulations cover a spectrum of inverse couplings, ranging from $\beta=2.0-2.8$, corresponding to lattice spacings from extremely coarse to very fine. We use a Langevin algorithm with an improved update step \cite{Ukawa:1985hr} with step size $\Delta \tau=10^{-3}$ to generate gauge configurations starting from homogeneous configurations of identity matrices (cold start).

The thermalization of the system is monitored through the evolution of the Polyakov loop expectation value, which approaches zero in the confined phase at zero temperature. We estimate the thermalization time by observing the Polyakov loop's plateau formation. We then simulate for 10-100 times longer than the time it takes for the plateau to appear, ensuring the thermalization of all infrared modes. Once thermalization is achieved, configurations are recorded with a distance of multiple autocorrelation times of the Polyakov loop. In this way, we record 100 independent sample configurations for each value of the inverse coupling and bare anisotropy. We emphasize that the flow time step has to be chosen carefully as the flow anisotropy enters quadratically in the discrete drift term of the flow equation Eq.\ \eqref{eq:flow_drift} and therefore can lead to numerical artifacts when chosen too large. 
More specifically, we have observed the emergence of a plateau for $\xi_{\mathrm{phys}}$ at values $\xi_{\mathrm{bare}} \gtrsim 6.5$ for a fixed flow time step $\Delta \theta_f = 0.01$. This numerical artifact can be removed by choosing $\Delta \theta_f = 0.01 / \xi_{f}^2$.
To ensure correctness, we check our results for sensitivity to Langevin and flow time steps as well as the thermalization time and the lattice volume. 

We flow the generated configurations using a third-order Runge-Kutta algorithm introduced in \cite{Luscher:2010iy} to solve the gradient flow equation Eq.\ \eqref{eq:gradient_flow}. Each configuration at some coupling $\beta$ and bare anisotropy $\xi_0$ is evolved with various flow anisotropies $\xi_f$, while the magnetic and electric contributions of the energy densities are tracked until at the condition \eqref{eq:B_flow_condition} is satisfied.
The flow trajectory of the ratio defined in Eq.\ \eqref{eq:flow_ratio} is shown in Fig.\ \ref{fig:flow_traj} for several flow anisotropies $\xi_f$ at the coupling $\beta=2.4$ and bare anisotropy $\xi_0=2$. The figure confirms that at a certain flow anisotropy, the flowed gauge configurations satisfy the equipartition relation. It is this flow anisotropy that approximates the physical anisotropy \cite{Borsanyi:2012zs}, as in Eq.~\eqref{eq:equi_anisotropy}.

Our main finding of this study is that the physical anisotropy $\xi_{\mathrm{phys}}$ depends linearly on $\xi_0$ to good accuracy. This is demonstrated in Fig.~\ref{fig:anisotropy} where we show $\xi_{\mathrm{phys}}$ obtained by the Wilson gradient flow procedure (dashed lines, error bars determined using the jackknife method) for $\xi_0 \leq 16$ with inverse coupling $\beta = 2.0$, $2.4$, $2.5$ and $2.6$. To make a more quantitative comparison, we employ a parametrization of the relative anisotropy $\eta=\xi_{\mathrm{phys}} / \xi_0$ that was motivated by perturbative arguments and previously used for SU(3) gauge theory in \cite{Klassen:1998ua},  
\begin{align} \label{eq:param} 
    \eta(\xi_{\mathrm{phys}}, \beta) = 1 + \hat{\eta}_1(\xi_{\mathrm{phys}}) \left(1-\frac{1}{\xi_{\mathrm{phys}}}\right) \frac{g^2}{2 N_c} \frac{1+ a_1 g^2}{1 + a_0 g^2}.
\end{align} 
Using $\xi_{\mathrm{phys}}=1-6$ and $\beta=5.5-24$, it was demonstrated there that $\hat{\eta}_1$ is approximately equal to 1 for SU(3), which we also assume for our tests in SU(2) theory. In Fig.~\ref{fig:anisotropy} we show the fitting function (solid lines) with the parameters 
\begin{align}
    a_0 = -0.403(5)\,, \qquad \quad a_1 =-0.481(1)\,,
    \label{eq:fit_parameters}
\end{align}
obtained using a least-squares fit to our data for $\beta = 2.0$ and $2.5$. Although the tested couplings may be in the non-perturbative regime \cite{Hirakida:2018uoy}, we observe a remarkably good description of our data. We note that this parametrization with the stated fitting parameters may receive corrections at other couplings. 

From perturbation theory, it is expected that the slope of the physical anisotropy converges to unity. In the previous study conducted in \cite{Klassen:1998ua} it was shown that this limit is approached monotonically for $\beta \to \infty$ for SU(3). Based on our simulation data for SU(2) in Fig.~\ref{fig:anisotropy} as well as their parametrization \eqref{eq:param} with \eqref{eq:fit_parameters}, we find that unity is approached non-monotonically. This observation is not necessarily a contradiction with previous studies as we simulate a different gauge theory. Nevertheless, this curious finding needs to be studied further in order to be confirmed. We note that for larger $\beta$, i.e.~closer to the perturbative regime, the required lattice sizes grow drastically due to finite volume effects, which makes such studies computationally demanding.

\section{Conclusion}
    We have studied the renormalization of SU(2) Euclidean lattice gauge theory. In particular, we simulated Yang-Mills theory on an anisotropic discretized lattice and used the Wilson gradient flow technique to obtain a scale as well as the physical anisotropy. The latter determines the relation between temporal and spatial lattice spacings in physical units.

    In all of our simulations, we have found an approximately linear relation between the physical and bare anisotropies. Motivated by a previous study for SU(3) gauge theory, we additionally fitted our results using a parametrization function with only two parameters. This function allows us to reproduce our data remarkably well for a wide range of anisotropies and several couplings corresponding to fine and very coarse lattices. More simulations with different couplings may further improve the accuracy of the parametrization of the data. 
    
    The observed almost linear dependence on the bare anisotropy and, in particular, this parametrization may enable an a priori tuning of the anisotropy. This may significantly reduce the computational effort of future SU(2) lattice calculations, especially when conducting the continuum limit for anisotropic lattices. Furthermore, this scale setting program is useful for modern real-time simulation methods: the stability of Complex Langevin on complex and real-time contours can be significantly improved by simulating on anisotropic lattices.

\begin{acknowledgments}
    The authors thank S.~Hands  and R.~Bignell for the discussion and many suggestions in the poster session of the Lattice 2023 conference. This research was partially funded by the Austrian Science Fund (FWF) under the projects P~34455 and P~36875. D.M.~acknowledges additional support from project P~34764. The computational results presented have been achieved using the Vienna Scientific Cluster (VSC).
  \end{acknowledgments}

\bibliographystyle{JHEP}
\bibliography{ref}

\end{document}